# Universality in short-range Ising spin glasses

E. Nogueira Jr.
*Instituto de Física,*
*Universidade Federal da Bahia,*
*Campus Universitário de Ondina,*
*40210-340, Salvador, Bahia, Brazil*

S. Coutinho
*Laboratório de Física Teórica e Computacional,*
*Universidade Federal de Pernambuco,*
*50670-901, Recife, Pernambuco, Brazil*

F. D. Nobre
*Departamento de Física Teórica e Experimental,*
*Universidade Federal do Rio Grande do Norte,*
*Campus Universitário,*
*CP 1641, 59072-970, Natal, Rio Grande do Norte, Brazil.*

E. M. F. Curado*
*Centro Brasileiro de Pesquisas Físicas,*
*Rua Xavier Sigaud, 150,*
*22290-180, Rio de Janeiro, Brazil,*
(August 11, 2018)

The role of the distribution of coupling constants on the critical exponents of the short-range Ising spin-glass model is investigated via real space renormalization group. A saddle-point spin glass critical point characterized by a fixed-point distribution is found in an appropriated parameter space. The critical exponents $\beta$ and $\nu$ are directly estimated from the data of the local Edwards-Anderson order parameters for the model defined on a diamond hierarchical lattice of fractal dimension $d_f = 3$. Four distinct initial distributions of coupling constants (Gaussian, bimodal, uniform and exponential) are considered; the results clearly indicate a universal behavior.

The nature of the spin-glass (SG) transition on real systems remains a subject of controversies. There is no definite agreement among theories, numerical simulations and experimental results, about the standard critical exponents governing the transition, as it occurs for pure and disordered spin systems without frustration. Different apparatus used to measure the non-linear susceptibilities [1,2], distinct numerical procedures used in computational simulations [3–5] and the lack of an exact renormalization-group scheme to treat short-range spin-glass models in finite dimensions ($d > 1$), produce a spread in the values of the critical exponents, leading some authors to claim the absence of universality in the critical behavior of spin glasses [6]. The most popular renormalization-group method used to investigate spin-glass models on finite hypercubic lattices [7] has been the so-called Migdal-Kadanoff scheme [8], which was proven to be exact, for pure systems, on a class of diamond hierarchical lattices (DHL) [9]. Such a technique has been valuable in the study of short-range spin glasses [7,10–14] and was used recently to investigate the nature of the low temperature phase for an Ising SG on a DHL [14]. Usually, to treat the spin-glass problem within a renormalization-group (RG) approach one has to choose arbitrarily an initial distribution of exchange coupling constants. In the standard approach [7], the parameter space of the renormalization flow is one-dimensional and the corresponding renormalized quantity is the mean-square-root deviation of $K_{ij} = J_{ij}/k_B T$, where $\{J_{ij}\}$ are the random quenched nearest-neighbour coupling constants, $k_B$ is the Boltzmann constant and $T$ is the temperature. For different chosen initial distributions $P(\{K_{ij}\})$, different critical temperatures and sets of critical exponents have been found either using the RG scheme or performing numerical simulations [6,10].

In the present work, the role of the initial distribution on the critical properties of the Ising spin-glass model is investigated. We consider the short-range Ising spin glass defined on the DHL with fractal dimension $d_f = 3$, described by the Hamiltonian,

$$H = -\sum_{<i,j>} J_{ij}\sigma_i\sigma_j, \qquad (1)$$

where the $\sigma$'s represent the Ising variables assigned to the lattice sites and the sum is restricted to the nearest-neighbour sites. The evolution of the distributions, under the renormalization process, is anal-ysed in an appropriated higher-dimensional parameter space. A two-dimensional projection of this space is the plane $< tanh^2(J_{ij}/k_BT) >$ versus $k_BT/ < J_{ij}^2 >^{1/2}$. In this representation, each symmetric distribution $P(\{J_{ij}\})$ is assigned to a curve defined by $F_P(T) =< tanh^2(J_{ij}/k_BT) >_P$, where $< ... >_P$ means an average over the distribution [11]. This parameter space is separated in two regions, each one governed by its respective attractor, the spin-glass $(1,0)$ and paramagnetic $(0,\infty)$ stable fixed points. The frontier separating these regions is defined by the locus of $F_P(T_c^P)$, $T_c^P$ being the critical temperature, above (below) which the flow evolves to the paramagnetic (spin glass) stable fixed point. Within this frontier, one finds the saddle-point spin-glass critical point, characterizing the "fixed-point" distribution $P^*(\{K_{ij}\})$, which should remain invariant under renormalization. In Figure 1, we display a realization of the flow diagram in the vicinity of the saddle-point spin-glass critical point, considering four initial distinct symmetric distributions, namely the *Gaussian, bimodal, exponential* and *uniform* distributions,

$$P(J_{i,j}) = \frac{1}{\sqrt{2\pi}} \exp(-\frac{1}{2}J_{i,j}^2),$$
$$P(J_{i,j}) = \frac{1}{2}\left[\delta(J_{i,j} - 1) + \delta(J_{i,j} + 1)\right],$$
$$P(J_{i,j}) = \frac{1}{\sqrt{2}} \exp(-\sqrt{2}|J_{i,j}|), \quad (2)$$
$$P(J_{i,j}) = \begin{cases} \frac{1}{2\sqrt{3}} & if \ -\sqrt{3} \leq J_{i,j} \leq \sqrt{3} \\ 0 & (otherwise). \end{cases}$$

From the data of this plot we estimate the critical temperatures $T_c^P$ corresponding to each distribution, recovering some previous results [7], as well as the critical temperature $T_c^*$ corresponding to the "fixed-point" distribution as shown in Table 1. We notice that the positions of the $F_P(T)$ curves in Figure 1 follow a decrease of the kurtosis $\kappa =< J_{ij}^4 > / < J_{ij}^2 >^2$ of the corresponding distributions, as we go from the left to the right. We have estimated the kurtosis of the "fixed-point" distribution, obtaining $\kappa \simeq 3.30$, i.e. is close to the Gaussian one.

In a previous work [10], we have made a direct estimate of the critical exponents associated with the order parameter ($\beta$ exponent) and the correlation length ($\nu$ exponent) for the corresponding above-mentioned initial distributions. In that approach the exponents were computed from the set of numerical values of the local Edwards-Anderson (EA) order parameter, obtained through an exact recursive procedure developed by the authors [12]. Within this methodology, the values of the local EA order parameter, for the sites introduced at each hierarchy, are calculated taking into account the random couplings generated by the corresponding renormalized distribution. Therefore, in the vicinity and below $T_c^P$, the values of the local EA order parameter of the last hierarchy were calculated with the coupling constants given by the initial distribution. Since, for the present DHL, $\frac{7}{8}$ of the total number of sites belong to the last generation, different values of $\beta$ appear, associated with the behavior of the system close to these critical points, suggesting a breaking of universality [10].

To overcome this apparent breaking of universality, herein we estimate the critical exponents by calculating the temperature dependent EA order parameter close to the saddle-point spin-glass fixed point, taking into account the "fixed-point" distribution $P^*(\{K_{ij}\})$. Since the exact analytical form of this distribution is unknown, these exponents can not be determined, unless one carefully probes the critical region. For that, we monitored numerically the evolution of the coupling constant distribution until it reaches the $F^*(T)$ curve at the vicinity of saddle-point spin-glass critical point. This is done by following the flow in the diagram $< tanh^2(J_{ij}/k_BT) >$ versus $k_BT/ < J_{ij}^2 >^{1/2}$ and counting the number of steps until the flow reaches the $F^*(T)$ curve. For the present DHL this behaviour occurs after the fourth renormalization step, as can be seen in Figure 1.

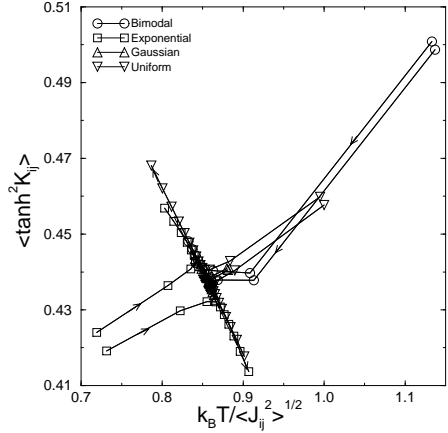

FIG. 1. The evolution of the probability distributions, under the renormalization process, in a two-dimensional space.

From this point on, we apply the method developed in [12], calculating the local magnetization of each lattice site and the EA order-parameter, varying the temperature and taking an average over many initial distribution realizations (samples). The critical exponent associated with the EA order parameter per spin is defined by $q_{EA} \sim [(T_c^* - T)/T_c^*]^\beta$ where $T_c^*$ is the spin-glass critical temperature corresponding to the saddle-point spin-glass critical point. Moreover, the correlation-length critical exponent can be also estimated by scaling the order pa-rameter at $T = T_c^*$, i.e., $Q_{EA} \sim M^{\beta/\nu}$ where $M = b^N + 1$, $b$ being the scaling factor (herein we restrict ourselves to $b = 2$) and $N$ the number of lattice hierarchies. We consider the four above-mentioned distributions of coupling constants as the initial ones, evaluating the corresponding $\beta$ critical exponents in each case, for lattices up to fifteen generations and 200 samples. To obtain the ratio $\beta/\nu$ we consider $N$ varying from 8 to 15 and the number of samples varying from 1000 (for smaller systems) to 300 (larger systems).

In Figure 2, we exhibit the plot of the EA order parameter per spin as a function of the temperature for each case. Note that all plots fall onto the same curve within the errors bars. One notices that larger error bars are found as one gets away from the critical temperature $T_c^*$, contrary to what happens in conventional numerical simulations. This is due to the fact that our sample averages are taken over different initial conditions (i.e., distinct pools of numbers associated to a given initial distribution [7]).

generated by the considered algorithm avoid the reach of the saturated limit. The corresponding points in the flow diagram, indicating where the order parameter was calculated, all fall onto the $F^*(T)$ curve characterizing the spin-glass "fixed-point" distribution, as shown in the inset of Figure 2.

In Figure 3, we exhibit the log-log plot of the data in Figure 2. The results from distinct initial distributions all coincide within the error bars, indicating a universal behaviour for the $\beta$ exponent. The scaling region used for the computation of the $\beta$ exponents was $ln[(T_c^* - T)/T_c^*]$ in the range from $-4.0$ to $-2.0$; small variations in the width of this scaling region (e.g., from $-4.0$ to $-1.5$) did not change significantly our estimates ($\beta$ remainig inside the error bars). A similar universal behavior is observed in Figure 4, where we exhibit the logarithm of $Q_{EA}$ as function of the hierarchical level $N$, at the critical temperature $T_c^*$, yielding the same $\beta/\nu$ ratio within the error bars.

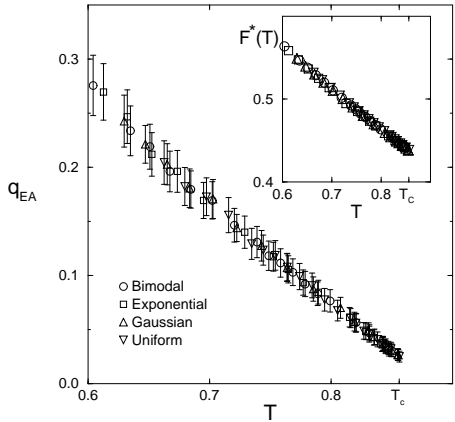

FIG. 2. The EA order parameter per spin as a function of the temperature (in units $k_B = 1$), in the neighbourhood of $T_c^*$. The inset shows points in the flow diagram were the values of $q_{EA}$ where calculated.

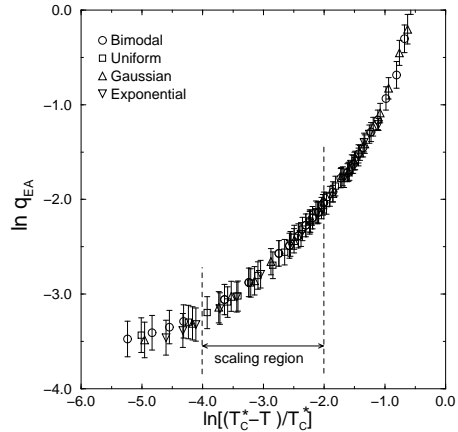

FIG. 3. The log-log plot of the EA order parameter per spin in terms of the distance from the critical temperature $T_c^*$. One clearly sees a universal behaviour within the error bars. The values of the $\beta$ exponents were computed from the points inside the scaling region shown.

Close to the critical point the distribution parameters (e.g., its mean-square deviation) vary slowly and so are not very sensitive to such initial conditions, whereas away from $T_c^*$, different sets of numbers may lead to RG trajectories far apart in the parameters space. For temperatures very close to the critical point, finite-size effects occur, preventing the vanishing of the order parameter at $T_c^*$, while for low temperatures, numerical difficulties

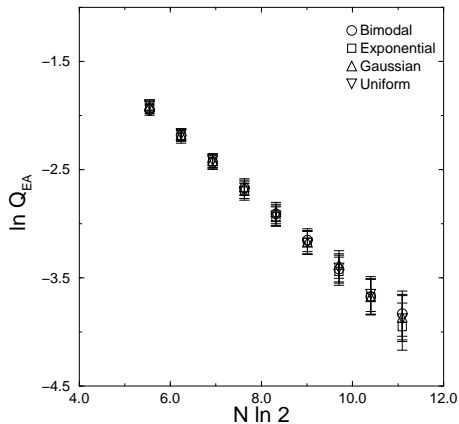

FIG. 4. The logarithm of the EA order parameter per spin as a function of the hierarchical level, at the universal spin-glass critical temperature $T_c^*$.

In Table 1, we exhibit the corresponding values of $\beta$, $\beta/\nu$ and the estimated values of $\nu$, corresponding to each initial distribution. We call the reader's for the fact that these "universal" values for the critical exponents are very close to the ones obtained by considering the Gaussian distribution as the initial one within the approach of reference [10]. This is an expected result since the "fixed-point" distribution is very close to the Gaussian one [10,11]. This can also be seen in Figure 1 by noting that the point corresponding to the Gaussian distribution at $T_c^P \simeq 0.88$ is indeed very close to the curve $F^*(T)$.

TABLE I. The estimated values of the critical temperature $T_c^P$ for each initial distribution of coupling constants, the universal spin-glass critical temperature $T_c^*$, and the critical exponents $\beta$ and $\nu$. The columns are disposed according to decreasing kurtosis of the corresponding initial distribution.

|         | Exponential | Gaussian | Uniform | Bimodal |
|---------|-------------|----------|---------|---------|
| $T_c^P$ | 0.723 ±0.001 | 0.881 ±0.001 | 0.998 ±0.001 | 1.132 ±0.001 |
| $T_c^*$ | 0.854 ±0.002 | 0.856 ±0.002 | 0.856 ±0.001 | 0.854 ±0.002 |
| $\beta$ | 0.63 ±0.03 | 0.63 ±0.03 | 0.63 ±0.03 | 0.63 ±0.03 |
| $\beta/\nu$ | 0.36 ±0.01 | 0.36 ±0.01 | 0.36 ±0.01 | 0.35 ±0.01 |
| $\nu$ | 1.8 ±0.1 | 1.8 ±0.1 | 1.8 ±0.1 | 1.8 ±0.1 |
| $\kappa$ | 6.0 | 3.0 | 1.8 | 1.0 |

It should be stressed, as discussed in [10], that the agreement of the present results with those obtained from other methods, developed for the cubic lattice (e.g., extensive Monte Carlo simulations [4,5], high-temperature series expansions [15]) is very impressive, taking into account that the former are, in principle, valid for a $d_f = 3$ diamond hierarchical lattice. It should be mentioned that, as far as the hierarchical lattice is concerned, the only approximations involved in the present approach are the finite sizes, and the number of samples investigated. A further analysis of the good performance of the short-range Ising SG model defined on a DHL, with scaling factor two and fractal dimension three, was presented in ref. [13].

In conclusion, we have shown that, as far as the Migdal-Kadanoff renormalization group scheme is concerned, the critical behavior of the short-range Ising SG model is governed by a unique saddle-point spin-glass fixed point, characterized by a fixed-point distribution of coupling constants. Furthermore, its critical exponents are universal and independent of the particular initial distribution of coupling constants employed. Whether this scenario will persist for more sophisticated approaches remains an open question, now under study.

### ACKNOWLEDGMENTS

This research was partially supported by CNPq, FINEP and CAPES (Brazilian granting agencies). One of us (S.C.) is particularly grateful to FACEPE for the financial support under the grant ACE 0456-1.05/98.

* Also at International Centre of Condensed Matter Physics and Departamento de Física, Universidade de Brasília, C. P. 04667, 70919-970, Brasília, Brazil.
[1] K. H. Fischer and J. A. Hertz, *Spin Glasses* (Cambridge University Press, Cambridge, 1991).
[2] A. V. Deryabin, V. K. Kazantsev and I. V. Zakharov, Sov. Phys. Solid State **30**, 129 (1988).
[3] R. N. Bhatt and A. P. Young, Phys. Rev. Lett. **54** 924; (1985); A. T. Ogielski and I. Morgenstern, Phys. Rev. Lett. **54**, 928 (1985).
[4] R. N. Bhatt and A. P. Young, Phys. Rev. B **37**, 5606 (1988).
[5] N. Kawashima and A. P. Young, Phys. Rev. B **53**, R484 (1996).
[6] L. W. Bernardi, and I. A. Campbell, Phys. Rev. B **49**, 728(1994); **52**, 12501 (1995); Europhys. Lett. **26**, 147 (1994); L. W. Bernardi, S. Prakash and I. A. Campbell, Phys. Rev. Lett. **77**, 2798 (1996).
[7] B. W. Southern and A. P. Young, J. Phys. C **10**, 2179 (1977).

[8] A. A. Migdal, Sov. Phys. JETP **42**, 743 (1975); L. P. Kadanoff, Ann. Phys. (New York) **91**, 226 (1975).

[9] For a recent review on hierarchical lattices, see C. Tsallis and A. C. N. Magalhães, Phys. Rep. **268**, 305 (1996).

[10] E. Nogueira-Jr., S. Coutinho, F. D. Nobre and E. M. F. Curado, Physica A **257**, 365 (1998).

[11] E. M. F. Curado and J-L. Meunier, Physica A **149**, 164 (1988).

[12] E. Nogueira-Jr, S. Coutinho, F. D. Nobre, E. M. F. Curado and J. R. L. de Almeida, Phys. Rev. E **55**, 3934 (1997).

[13] S. Prakash and I. A. Campbell, Physica A **235**, 507 (1997).

[14] M. A. Moore, H. Bokil and B. Drossel, Phys. Rev. Lett. **81**, 4252 (1998).

[15] L. Klein, J. Adler, A. Aharony, A. B. Harris and Y. Meir, Phys. Rev. B **32**, 11249 (1991).